\documentclass{ws-ijmpa}
\usepackage[super,compress]{cite}
\usepackage{graphicx}

\def\be{\begin{equation}}
\def\ee{\end{equation}}
\def\ba{\begin{eqnarray}}
\def\ea{\end{eqnarray}}
\def\order#1{{\mathcal O}\left(#1\right)}

\def\gsim{\mathrel{\raise.3ex\hbox{$>$\kern-.75em\lower1ex\hbox{$\sim$}}}}
\def\lsim{\mathrel{\raise.3ex\hbox{$<$\kern-.75em\lower1ex\hbox{$\sim$}}}}

\begin{document}
\markboth{A.B. Arbuzov, T.V. Kopylova, I.K. Sklyarov}{On spin asymmetry in muon and tau decays}

%
\catchline{}{}{}{}{}
%

\title{On spin asymmetry in muon and tau decays}

\author{A.B. Arbuzov$^{1,2,\dagger,}$\footnote{Corresponding author.}\ ,\ \ \ 
T.V. Kopylova$^{2}$\ and I.K. Sklyarov$^{1,2}$}

\address{$^1$Bogoliubov Laboratory of Theoretical Physics,
Joint Institute for Nuclear Research,\\ Dubna, 141980, Russia \\
$^2$Dubna State University, Universitetskaya str. 19, Dubna, 141980, Russia
\\
$^\dagger$arbuzov@theor.jinr.ru}

\maketitle


\begin{abstract}
The angular asymmetry in decays of polarized muons and tau
leptons is discussed. Both the standard $V-A$ Fermi model and 
the general parameterization via Michel parameters 
are considered. Numerical importance of contributions
suppressed by charged lepton mass ratio is underlined. 
Contribution of the second order QED correction is estimated
in the leading logarithm approximation.

\end{abstract}

\ccode{PACS numbers: 
13.35.Bv, 
13.35.Dx, 
12.15.Ji 
}


\section{Introduction}
\label{sect:Intro}

Extremely accurate experiments on the muon decay 
lifetime~\cite{Tishchenko:2012ie,Barczyk:2007hp} 
give the value of the Fermi coupling constant $G_{\mathrm{Fermi}}$
with the precision of about 0.5~ppm. That provides normalization 
for evaluation of many observables in electroweak 
physics starting from differential distributions in muon decays up to
the $W$ boson mass measurements~\cite{Alioli:2016fum}, high-precision
tests of the Standard Model (SM), and new-physics searches.
New experiments with high statistics on tau lepton decays provide
a good possibility to test the lepton universality hypothesis
with increased accuracy.   

Effects suppressed by the electron to muon mass ratio in the muon 
decay spectrum were computed within the standard $V-A$ Fermi model in 
the $\order{\alpha}$ order~\cite{Arbuzov:2001ui}
and in the $\order{\alpha^2}$ order~\cite{Pak:2008qt} 
approximations in perturbative QED. 
Recently in Ref.~\cite{Arbuzov:2016ywn} we also considered the mass 
effects in radiative muon and tau lepton decays within 
the model-independent approach. 

Here we would like to continue discussion of the spin asymmetry 
in muon decay started in Ref.~\cite{Caola:2014daa}. This quantity 
is an inclusive observable and can be potentially measured
with a high accuracy. We will include in consideration
leptonic decay modes of tau lepton. Besides the pure $V-A$ case, the asymmetry
will be treated in the model-independent approach with the help of 
the Michel parameters. 
Special attention will be payed to the terms suppressed by the
ratio of the produced charged lepton mass to the decaying one.

At the Born level the spin asymmetry can be defined as an explicitly 
inclusive observable. Meanwhile taking into account higher order
corrections due to emission of real photons and light lepton pairs
makes the situation more complicated or even ambiguous when several 
charged leptons appear in the final state. In order to compute the  
higher order contributions one should rely on a concrete experimental 
set-up. Both theoretical and experimental definitions of the asymmetry
should be infrared safe, {\it i.e. } be numerically stable with respect
to variations of soft and collinear radiation. Two choices of such 
definitions will be discussed below.
 
The paper is organized as follows. The next section contains 
the notation and preliminary remarks. Sect.~\ref{Sect:spin_asym} is devoted
to the spin asymmetry treated in the two approaches mentioned above.
Numerical estimates of the effects due to radiative corrections and
the terms suppressed by mass ratio are also given there.
Sect.~\ref{Sect:concl} contains concluding remarks.

\section{Preliminaries and notation}

Within the SM, muon decays are described by interactions of vector currents
formed by left fermions. Meanwhile, many models beyond the SM predict 
contributions of other kinds. Since the energy scale of new physics is 
(most likely) higher than the electroweak scale, the corresponding 
contributions can be parameterized by four-fermions interactions with 
different currents and coupling constants, see {\it e.g.}~\cite{Olive:2016xmw}.

Let us consider the general decay $L^-\rightarrow l^- \bar{\nu}_l \nu_L$ 
in the rest frame of the heavy lepton ($L$ is either $\mu$ or $\tau$,
and the light lepton is either electron or muon). 
The differential distribution in the energy and angle of the final state
charged lepton can be described with the help of the Michel parameters 
$\rho_L$, $\xi_L$, $\delta_L$, $\eta_L$~\cite{Michel:1949qe,Bouchiat:1957zz,Kinoshita:1957zz,Fetscher:1986uj}:
\begin{eqnarray} \label{dGamma}
\frac{d\Gamma (L^-\rightarrow l^- \nu \bar{\nu})}{dx\, d\cos\theta_l}&=&
  \frac{m_L^5 (1+r^2)^4}{32\pi^3}\sqrt{x^2-x_0^2}G_{0,L}^2
  \left[F_{IS}(x)-F_{AS}(x){\cal{P}}_L \cos\theta_l \right],\nonumber \\
    F_{IS}(x)&=&x(1-x)N+\frac{2N\rho_L}{9}(4x^2-3x-x_0^2)
  +N\eta_L x_0 (1-x),\nonumber \\
  F_{AS}(x)&=&
  \frac{N\xi_L\sqrt{x^2-x_0^2}}{3}
  \left\{ 1-x +
  \frac{2\delta_L(4x-4+\sqrt{1-x_0^2})}{3}\right\},
\end{eqnarray}
where ${\cal P}_L$ is the polarization degree of the initial lepton. 
In order to account possible violation of the lepton universality, 
we introduced the lepton index $L$ for the effective Fermi constant 
$G_{0,L}$ and for the Michel parameters\footnote{In general, the tau lepton Michel parameters might also depend on the light lepton choice.}.
Angle $\theta_l$ is chosen between the $L^-$ spin and 
the final $l^-$ momentum. The energy fraction of the produced charged lepton
is denoted as $x=E_l/E_l^{\mathrm{max}}$. Its minimal value is
$x_0=2r/(1+r^2)$, where $r=m_l/m_L$; $m_l$ and $m_L$ are 
the masses of the decaying and produced charged leptons, respectively.
The maximal lepton energy is $E_l^{\mathrm{max}} = m_L(1+r^2)/2$.
The constant $N$ and its products with the Michel parameters 
$N\rho_L$, $N\xi_L$, $N\xi_L\delta_L$, and $N\eta_L$ are certain bilinear
combinations of the four-fermion coupling constants~\cite{Olive:2016xmw}. 
Their values are defined from the fits of experimental data on 
differential and integrated decay distributions (or taken from the SM).

\section{Muon decay spin asymmetry}
\label{Sect:spin_asym}

Let us define the spin asymmetry in the leptonic decays of muon 
and tau lepton as
\begin{eqnarray} \label{A_def}
A_L = \frac{1}{\Gamma_{L}}\int\limits_{x_0}^{1}dx 
\left\{ \int\limits_{0}^{1} - \int\limits_{-1}^{0} \right\}d\cos\theta_l \cdot
\frac{d\Gamma (L^-\rightarrow l^- \nu \bar{\nu})}{dx\, d\cos\theta_l},
\end{eqnarray}
where $\Gamma_{L}$ is the total decay width, 
\begin{eqnarray}
\Gamma_{L} = \int\limits_{x_0}^{1}dx \int_{-1}^{1} d\cos\theta_l 
\frac{d\Gamma (L^-\rightarrow l^- \nu \bar{\nu})}{dx\, d\cos\theta_l}.
\end{eqnarray}
Our definition of the asymmetry differs by factor 2 from the
one adapted in Ref.~\cite{Caola:2014daa} where instead of integration
over the angle just the difference between the values at $\cos\theta_l=+1$
and $\cos\theta_l=-1$ was used. We also suggest to normalize the asymmetry
by the total decay width instead of the tree level one used in 
Ref.~\cite{Caola:2014daa}.

\subsection{Spin asymmetry in the model-independent approach}

In the model-independent approach at the Born level the asymmetry 
is equal to the integral of function $F_{AS}(x)$ from Eq.~(\ref{dGamma}):
\begin{eqnarray} \label{A0M}
A_0 &=& - \int_{x_0}^1 F_{AS}(x) dx= -\int\limits_{x_0}^1 dx
\frac{1}{3}N\xi_L(1-x)\sqrt{x^2-x_0^2} 
\nonumber \\
&-& \int\limits_{x_0}^1 dx \frac{1}{3}{N\xi_L \delta_L\sqrt{x^2-x_0^2}}
{(4x-4+\sqrt{1-x_0^2})} 
\nonumber \\
&=& - \frac{1}{6}\left\{ N\xi_L g_{\xi}(r) 
+ N\xi_L \delta_L g_{\xi \delta}(r)\right\},
\end{eqnarray}
where 100\% polarization $({\cal{P}}_L=1)$ is assumed and
\begin{eqnarray}
g_{\xi}(r) &=& 1 - 20r^2 + 64r^3 - 90r^4 + 64r^5 -20r^6 +r^8,
\nonumber  \\
g_{\xi \delta}(r) &=& \frac{80}{3}r^2 - 128r^3 + 240r^4 
- \frac{640}{3}r^5 + 80r^6 - \frac{16}{3}r^8.
\end{eqnarray}
Note that the coefficients in front of the terms proportional
to $r^2$ are numerically large. So these terms might be relevant
for the analysis of the asymmetry in the decay $\tau\to\mu\bar{\nu}_\mu\nu_\tau$
for which $r=m_\mu/m_\tau\approx 0.06$.

\subsection{Asymmetry in the $V-A$ case}

One can see that in the pure $V-A$ case where $N\xi_L=1$
and $N\xi_L \delta_L=3/4$ for the Born-level asymmetry~(\ref{A0M})
we reproduce the known result, see
{\it e.g.}~\cite{Arbuzov:2001ui}, 
\begin{eqnarray} \label{eq:g0}
A_0(r) &=& - \frac{1}{6}g_0(r)\equiv A_0(0)g_0(r),
\nonumber  \\
g_0(r) &=& 1 - 32r^3 + 90r^4 - 96r^5 + 40r^6 - 3r^8.
\end{eqnarray}
Interestingly, the terms proportional to $r^2$ are canceled out
in the sum $g_{\xi}(r) + \frac{3}{4}g_{\xi \delta}(r)$.

Within the perturbative QED, the spin asymmetry can be presented 
in the form
\ba 
A = A_0(0)\left[g_0(r) + ag_1(r) + a^2g_2(r) + \order{a^3} \right], 
\ea 
where $a\equiv\alpha/\pi$,
and $\alpha\approx 1/137.036$ is the fine structure constant. 
The explicit expression for function 
$G_1(r)=4A_0g_1(r)$ can be found in Ref.~\cite{Arbuzov:2001ui},
while here we will use its expansion in $r$:
\ba \label{eq:g1}
g_1(r) &=& \frac{617}{72} - 7\zeta(2)
+ 4r + r^2\biggl( 48 - 24\zeta(2) - 2\ln r \biggr)
\nonumber \\
&+& r^3 \biggl( - \frac{284}{9} - \frac{112}{3}\ln r \biggr)
+ \order{r^4}.
\ea

As one can see, the one-loop QED correction to the asymmetry is free
from mass singularities, {\it i.e.} it is finite in the limit $r\to 0$.
That is in accord with the Kinoshita--Lee--Nauenberg theorem. 
Contributions proportional to the logarithms of mass ratios, 
like $\ln(m_\mu^2/m_e^2)\approx 10.66$, 
are canceled out in sufficiently inclusive observables, including
the total decay width and the spin asymmetry. Nevertheless, 
the dependence on the mass ratio logarithms remains in the effect
of the fine structure constant running which gives 
$\alpha(0)\approx 1/137.0$ $\to$ $\alpha(m_\mu)\approx 1/135.9$
and $\alpha(m_\tau)\approx 1/134.4$. 

The $\order{\alpha^2}$ contribution to the asymmetry was computed
in Ref.~\cite{Caola:2014daa} under a quite specific assumption of
experimental setup. Namely, it was assumed that components of an
electron-positron pair emitted at a small angle with respect to the
final {\em primary} electron are recombined with the parent particle, 
{\it i.e.} a calorimetric event selection is applied. 
A parameter to define the QED jet resolution was introduced. 
The QCD-like construction was motivated 
by the necessity to have an infrared safe definition of the asymmetry.

The suggested definition is infrared safe only under conditions
of the calorimetric event selection which is typical
for hadron jet registration at high-energy colliders.
But the definition has nothing to do with the typical experimental
set-ups used in muon decay spectrum measurements.
Indeed, most such experiments exploit the possibility 
to define the electron momentum from its curved spiral track 
in a magnetic field. In particular it is so in the TWIST 
experiment~\cite{Rodning:2001js,MacDonald:2008xf}. 
Obviously, a sufficiently strong  magnetic field
destroys the structure of any QED jet (consisting either of one electron
plus some photons or of several charged particles, like $e^+e^-e^-$),
and the observed trajectories of all charged particles become separated 
from each other. In this case the treatment of events with
several charged particles suggested in Ref.~\cite{Caola:2014daa}
is not infrared safe. Moreover, treatment of the spin asymmetry 
in events with observation of several charged tracks becomes ambiguous.

On the other hand, in experiments on tau lepton decays the momentum
of the final charged lepton is typically much higher than in the case
of muon decays, especially if the tau lepton decays in flight
like at the LHC. So the set-up considered in in Ref.~\cite{Caola:2014daa}
might be relevant for the experiments on leptonic tau decays
in flight. 

In order to estimate the $\order{\alpha^2}$ contribution in the case
when the presence of a strong magnetic field destroys the QED jet
structure, we suggest to use the following (simplified) set-up.
Let us assume that if the detector sees more than one charged particle track,
such an event is just dropped. It this case decays accompanied by production
of a real $e^+e^-$ pair will be not accounted in the integrated
decay asymmetry. But if the components of the pair have small energies
they will not be detected at all. To simulate this situation, let us
impose the simple cut-off on the total energy of the produced pair:
events with real pairs with energy exceeding $\Delta m_L/2$ are
dropped. Here $m_L$ is the mass of the decaying lepton and $\Delta\ll 1$
is a small parameter. The energy of the detected charged lepton
(which exhibits the angular asymmetry) should be above the cut-off $\Delta m_L/2$.
This definition of the asymmetry is obviously infrared safe because the 
domain of soft radiation is integrated out and the events with collinear 
pair emission are cut off. Numerical results below will be given for
$\Delta=0.1$ which corresponds to the maximal $e^+e^-$ pair energy 
$E^{\mathrm{max}}\sim 10$~MeV in the muon decay and 
$E^{\mathrm{max}}\sim 90$~MeV in the tau lepton decays 
$\tau\to e\bar{\nu}_e\nu_\tau e^+e^-$ and 
$\tau\to \mu\bar{\nu}_\mu\nu_\tau  e^+e^-$. 
For this event selection procedure we will have contributions
of the order $\order{\alpha^2\ln^2(m_L^2/m_l^2)}$ which are numerically
dominant at the two-loop level~\cite{Arbuzov:2002pp,Arbuzov:2002cn,Arbuzov:2002rp}. 
Moreover, the pair contribution will be also enhanced by the logarithm 
of the parameter $\Delta$. 
The described event selection removes the ambiguity in the treatment 
of events with two electrons in the final state. 
This also means that the quantum interference of these electrons is 
excluded since the energy domains of the primary electron and 
the secondary one (from the emitted pair) do not overlap.  

There is also the so-called singlet channel kinematical situation
when the primary electron and the secondary positron are soft, 
while the secondary electron has a large energy and it is detected instead 
of the primary one. The corresponding contribution can be estimated
by looking at the integral of the singlet part of the electron structure
function ${\mathcal D}^{\mathrm S}_{e^+e^-}(x)$, see {\it e.g.}~\cite{Arbuzov:2002pp}, 
over the interval $[1-\Delta,1]$.
It is also enhanced by the square of the large logarithm but it is suppressed 
by the second power of the small parameter $\Delta$. That allows to neglect
the singlet-channel contribution. 
 
One should note that in Eq.~(\ref{A_def}) we defined the asymmetry as 
the integral over the total range of the observed electron energy. While
as we discussed above the low-energy region (below $\Delta m_L/2$) might 
be not accessed experimentally. Here we will use the approximation where
this region for the primary electron is not dropped. This would not much
affect our estimates of the higher-order and mass effects since the electron
spectrum at small energy fraction values $x\equiv 2E_e/m_\mu$ behaves  
as $\sim x^2$. But in a realistic application one should take into account
the threshold of electron registration.

In the leading logarithmic approximation under the discussed experimental
conditions, the emission of an extra (virtual or real soft) $e^+e^-$ pair gives
\ba
g_2^{LLA}(r,\Delta) = \frac{1}{4}\ln^2(r^2)\left(2\ln\Delta+\frac{3}{2}\right).
\ea
Note that for the decay $\tau\to \mu e^+e^-\bar{\nu}_\mu\nu_\tau$ 
there are two types of mass singularities, that leads to the
substitution\footnote{This substitution was used to get the number
for $g_2^{LLA}$ in the last column in Table~\ref{Table:g}.}
$\ln^2(r^2)\to\ln(m_\mu^2/m_\tau^2)\cdot\ln(m_e^2/m_\tau^2)$. 
The NNLO result for virtual and soft $e^+e^-$ pair corrections 
to the spin asymmetry in muon decay was presented in Ref.~\cite{Arbuzov:2003ce}.

Numerical results for $g_0$, $g_1$, and $g_2^{LLA}$ are presented in
Table~\ref{Table:g}. 
For tau lepton decays we took into account only the electron-positron
pair contribution since its impact is enhanced by the large 
logarithm $\ln(m_\tau^2/m_e^2)\sim 16.3$ (even squared for 
$\tau\to 2e\bar{e}\bar{\nu}_e\nu_\tau$) while the logarithm with the 
muon mass is considerably smaller, $\ln(m_\tau^2/m_\mu^2)\sim 5.6$.

\begin{table}[ph]
\tbl{Coefficients $g_0$, $g_1$, and $g_{2}^{\mathrm{LLA}}$ 
{\it vs} the mass ratio $r$.}
{\begin{tabular}{@{}ccccc @{}} \toprule
\hline
 r         &\  0        &\ $m_e/m_\mu$ &\ $m_e/m_\tau$ &\ $m_\mu/m_\tau$ \\
 $g_{0}(r)$ &\ 1.000000  &\ 0.999996   &\ 1.000000   &\ 0.994327 \\ 
 $g_{1}(r)$ &\ -2.94509  &\ -2.92528   &\ -2.94394   &\ -2.64765 \\ 
 $g_{2}^{LLA}(r)$ 
           &\  ---      &\  -88.3       &\ -206.5     &\ -71.5   \\ \botrule
\end{tabular}\label{Table:g}}
\end{table}

We would like to underline that our estimates of the second order
contributions are performed only to have an idea about the size of the
effect. If the experimental uncertainty is of the same order or lower,
one should perform a Monte Carlo generation of decays with emission of
additional photons or electron-positron pairs and let the events pass through
a detector simulator. 

The numerical result for asymmetry in Ref.~\cite{Caola:2014daa} was given by
Eq.~(9) in the form
\ba \label{eq:Caola}
A = A_0\left[1 - 2.9451\; \bar{a} + 11.2(1)\, {\bar{a}}^2 \right], 
\ea 
where $\bar{a}\equiv\alpha_{\overline{\mathrm{MS}}}(m_\mu)/\pi$
and $\alpha_{\overline{\mathrm{MS}}}(m_\mu)\approx 1/135.9$ is the QED
coupling constant in the $\overline{\mathrm{MS}}$ scheme at the
muon mass scale. Calculations of the second order pair corrections in 
Ref.~\cite{Caola:2014daa} were also performed in the 
$\overline{\mathrm{MS}}$ scheme. Comparisons of the numerical results
for the radiative corrections to muon decay asymmetry given in Eq.~(\ref{eq:Caola})
and Table~\ref{Table:g} show a considerable difference in the $\mathcal{O}(\alpha^2)$
order. The difference consists of two parts: 
the one in the $g_2$ coefficient value and
the one given by the scheme change, {\it i.e.} 
$\bar{a}\cdot 2.9451 - (\alpha/\pi)g_1(0)$.
The main source of the deviation is due to the different treatment of 
the events with real $e^+e^-$ pair creation. 

Paper~\cite{Caola:2014daa} states that 
``the radiative corrections are more important than the electron mass effects.''
This statement is not fully correct. Obviously for the muon decay case
considered in Ref.~\cite{Caola:2014daa}, the $r^3$ mass corrections at the tree 
level, see Eq.~(\ref{eq:g0}), are small compared even to the $\order{\alpha^2}$
contribution of radiative corrections and one can safely take the limit 
$g_0(r)\to 1$.  
But the one-loop QED correction~(\ref{eq:g1}) contains the contribution
proportional to the first power of the mass ratio $r$, 
which makes it exactly of the same order as
the $\order{\alpha^2}$ contribution found in Ref.~\cite{Caola:2014daa}.
Namely, $4(m_e/m_\mu)\bar{a}\approx 8.3 \bar{a}^2$.
This contribution was missed in Eq.~(\ref{eq:Caola}).
It is interesting to note that such linear in mass terms are
not typical in the first order radiative corrections to isotropic
observables while they sometimes appear in asymmetries, see 
{\it e.g.} Ref.~\cite{Arbuzov:1991pr}.

\section{Conclusions}
\label{Sect:concl}

Taking into account the spin orientation of the initial particle
in decays of muon and tau lepton allows to get additional information
about the structure of weak interactions. High statistics on tau lepton
decays at several modern and future experiments admits high-precision
tests of the lepton universality hypothesis. 

In this paper we discussed the spin asymmetry in the muon decay
and the leptonic modes of tau decays. Explicit expressions for 
the asymmetry in model-independent description via the Michel parameters
at the Born level are derived with taking into account the terms
suppressed by the charged lepton mass ratio. These terms can be relevant
for future high-precision studies of the muon decay spectrum and especially
for experiments on the decay $\tau\to \mu {\bar\nu}_\mu \nu_\tau$.
We pointed out that the terms suppressed by the charged lepton mass ratio
are also numerically relevant in the contribution of the first order radiative 
corrections. This fact was discovered in Ref.~\cite{Arbuzov:2001ui} but 
it was missed in the earlier paper on muon decay asymmetry~\cite{Caola:2014daa}.
Radiative corrections and mass effects were considered for muon and tau lepton
decays in parallel. The second order QED corrections in the $V-A$ case were
estimated above for a simplified (but still reasonable) experimental set-up
with a cut-off on events with production of extra charged particles. 

The most accurate measurements of the differential distributions
in polarized muon decay were performed by the TWIST 
collaboration~\cite{MacDonald:2008xf,Bueno:2011fq,TWIST:2011aa}. The resulting 
uncertainty for the extracted Michel parameters reached the $10^{-4}$ order.
The TWIST experiment did not cover the full angular phase space and it was 
not suited to measure the decay asymmetry. There were also inclusive experiments
which measured the total muon decay width with the precision up to
0.5~ppm~\cite{Tishchenko:2012ie,Barczyk:2007hp}.
But the asymmetry was not measured there as well. 
We suggest to foresee the asymmetry measurement in future experiments 
on muon (and hopefully tau lepton) decays.
As discussed in the Introduction, a new more precise number for the 
muon lifetime can not serve to test the Standard Model, since the precision
of its theoretical prediction doesn't have such a high accuracy. 
But the asymmetry value is sensitive to the presence of non-standard
weak interactions and its measurement in future experiments potentially
can be performed with the accuracy better than $10^{-4}$ achieved
in the TWIST experiment.

In conclusion we would like to underline that the presented results can be
used as an estimate of higher order effects while to treat experimental data
on spin asymmetries one would need to perform Monte Carlo simulations. 

\section*{Acknowledgments} 
We are grateful to D.~Epifanov for useful discussions.

\end{document}